\documentclass[12pt]{book}
\def\f{{\rm p}}
\def\vd{{5}}
\def \vu {{(5)}\!}

\def \cu{ \rho_{\cal E}}
\def \cq{ q^{\cal E}}
\def \cp{ \pi^{\cal E}}

\def\be{\begin{equation}}
\def\ee{\end{equation}}
\def\bea{\begin{eqnarray}}
\def\eea{\end{eqnarray}}
\usepackage[dvips]{graphicx,color}
\usepackage{makeidx,physics,cosmology}
\usepackage{amsmath,amssymb}

\makeauthorindex \makeindex

\BookTitle{Frontier in Astroparticle Physics and Cosmology}
\CopyRight{Invited Talk at the 6th RESCEU Symposium, Nov. 4-7
2003, Tokyo, Japan.}

\begin{document}

\BookTitle{\itshape Frontier in Astroparticle Physics and
Cosmology}

\pagenumbering{arabic} \CopyRight{Invited Talk at the 6th RESCEU
Symposium, Nov. 4-7 2003, Tokyo, Japan.}
\chapter{Brane-world cosmological perturbations}

\author{
Roy MAARTENS\\
{\it Institute of Cosmology and Gravitation, University of
Portsmouth, Portsmouth PO1 2EG, UK}\\}

\AuthorContents{R.\ Maartens}

\AuthorIndex{Maartens}{R.}

\section*{Abstract}

Brane-world models provide a phenomenology that allows us to
explore the cosmological consequences of some M~theory ideas, and
at the same time to use precision cosmology as a test of these
ideas. In order to achieve this, we need to understand how
brane-world gravity affects cosmological dynamics and
perturbations. This introductory review describes the key features
of cosmological perturbations in a brane-world universe.

\section{Introduction}

Despite the tremendous successes of the concordance model (based
on general relativity and inflation), which can account for
high-precision cosmological observations, there remain deep
puzzles within this model. What is the fundamental theory
underlying inflation (or providing an alternative that matches its
successes)? What is the dark energy that appears to be dominating
the energy density of the universe and driving its late-time
acceleration, and how can it be explained by fundamental theory?
What is the dark matter that dominates over baryonic matter? These
unresolved questions at the core of the concordance model may be
an indication that high-precision cosmology is probing the limits
not only of particle physics, but also of general relativity. In
any event, it is important to pursue the cosmological implications
of quantum gravity theories -- and at the same time to subject
quantum gravity theories to the stringent tests following from
high-precision data.

The fully quantum regime entails the break-up of the space-time
continuum, but even when spacetime can be modelled as a continuum,
significant corrections to general relativity will arise at
energies below, but near, the fundamental scale. Traditionally,
the fundamental scale has been thought to be the Planck scale,
$M_\f\sim 10^{19}~$GeV. However, recent developments in M~theory,
a leading candidate quantum gravity theory~\cite{mtheory},
indicate that $M_\f$ may be an effective energy scale, with the
true fundamental scale being lower~\cite{add}. A key aspect of
M~theory is the need for extra spatial dimensions. If there are
$d$ extra (spatial) dimensions, with length scale $L$, then the
true fundamental scale $M_{4+d}$ is given by
 \be
M_\f^2 \sim M_{4+d}^{2+d}\,L^d\,.
 \ee
If $L\gg M_\f^{-1}$, then $M_{4+d}\ll M_\f$. Experiments in
colliders and table-top tests of gravitational force~\cite{cav}
imply the bounds $L\lesssim 0.1~$mm and $M_{4+d}\gtrsim 1~$TeV.

There are five distinct 1+9-dimensional superstring theories. In
the mid-90's duality transformations were found that relate these
superstring theories and the 1+10-dimensional supergravity theory,
leading to the conjecture that all of these theories arise as
different limits of a single theory -- M~theory. The 11th
dimension in M~theory is related to the string coupling strength;
the size of this dimension grows as the coupling becomes strong.
At low energies, M~theory can be approximated by 1+10-dimensional
supergravity. It was also discovered that p-branes, which are
extended objects of higher dimension than strings (1-branes), play
a fundamental role in the (non-perturbative) theory. Of particular
importance among p-branes are the D-branes, on which open strings
can end. Roughly speaking, open strings, which describe the
non-gravitational sector, are attached at their endpoints to
branes, while the closed strings of the gravitational sector can
move freely in the full spacetime (the ``bulk"). Classically, this
is realised via the localization of matter and radiation fields on
the brane, with gravity propagating in the bulk (see Fig.~1).

%%%%%%%%%%%%%%%%%%%%%%%%%%%
\begin{figure}[!bth]\label{brane}
\begin{center}
\includegraphics[height=3in,width=4in]{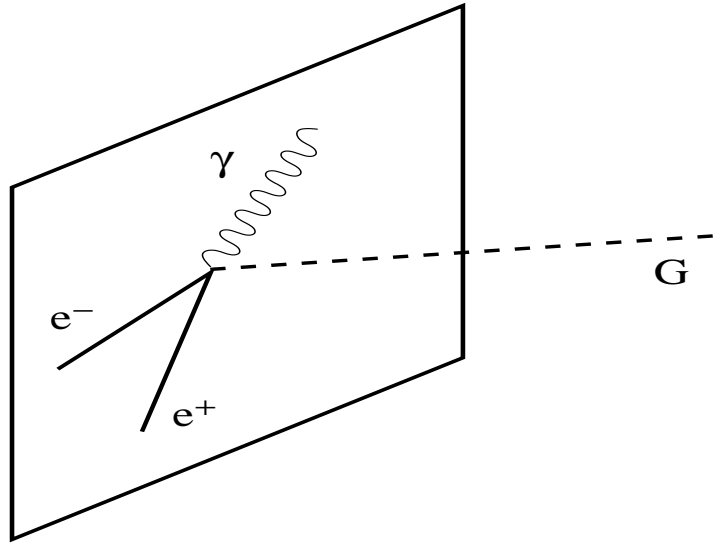}
\caption{Brane-world schematic: matter is confined to the brane,
while gravity propagates in the bulk (from~\cite{cav}).}
\end{center}
\end{figure}
%%%%%%%%%%%%%%%%%%%%%%%%%%%

In the Horava-Witten solution~\cite{hv}, gauge fields of the
standard model are confined on two 1+9-branes (or domain walls)
located at the end points of an $S^1/Z_2$ orbifold, i.e., a circle
folded on itself across a diameter. The 6 extra dimensions on the
branes are compactified on a very small scale, close to the
fundamental scale, and their effect on the dynamics is felt
through ``moduli" fields, i.e. 5D scalar fields. A 5D realization
of the Horava-Witten theory and the corresponding brane-world
cosmology is given in~\cite{low}. These solutions can be thought
of as effectively 5-dimensional, with an extra dimension that can
be large relative to the fundamental scale. They provide the basis
for the Randall-Sundrum type~1 models of 5-dimensional gravity,
with the bulk being a part of anti de Sitter spacetime
(AdS$_5$)~\cite{rs1}. The scalar degree of freedom describing the
inter-brane separation is known as the radion, and it requires a
stabilization mechanism if general relativity is to be recovered
at low energies.

As in the Horava-Witten solutions, the RS branes are
$Z_2$-symmetric (mirror symmetry), and have tensions (vacuum
energies) $\pm\sigma$, where
 \be\label{rst}
\sigma={3M_\f^2 \over 4\pi \ell^2}\,.
 \ee
Here $\ell$ is the curvature radius of the bulk, whose vacuum
energy (cosmological constant) is
 \be
 \Lambda_\vd=-{6\over \ell^2}\,.
 \ee
The single-brane Randall-Sundrum type~2 models~\cite{rs2} with
infinite extra dimension arise when the orbifold radius tends to
infinity. In this case, general relativity is recovered at low
energies; the weak-field gravitational potential is
 \be
\psi(r)={m\over r}\left(1+{2\ell^2 \over 3 r^2}\right)\,.
 \ee
The fundamental scale is given by
 \be\label{tt2}
M_\vd^3={M_\f^2 \over \ell}\,.
 \ee

\section{The brane observer's viewpoint}

Extra dimensions lead to new scalar, vector and tensor degrees of
freedom on the brane from the bulk graviton. In 5D, the spin-2
graviton is represented by a 4-transverse traceless metric
perturbation. In a suitable gauge, this contains: a 3-transverse
traceless perturbation -- the 4D spin-2 graviton (2
polarizations); a 3-transverse vector perturbation -- the 4D
spin-1 gravi-vector (or gravi-photon) (2 polarizations); and a
scalar perturbation -- the 4D spin-0 gravi-scalar (1
polarization). These modes of the 5D graviton are massive from the
brane viewpoint (essentially since the projection onto the brane
of the null 5D momentum is a timelike momentum on the brane). They
are known as Kaluza-Klein (KK) modes. The standard 4D graviton
corresponds to the massless zero-mode of the spin-2 part. In RS1,
the tower of massive modes is discrete, while in RS2 it is
continuous.

The novel feature of the RS models compared to previous
higher-dimensional models is that the observable 3 dimensions are
protected from the large extra dimension (at low energies) by
curvature rather than straightforward compactification. The RS
brane-worlds and their generalizations (to include matter on the
brane, scalar fields in the bulk, etc.) provide phenomenological
models that reflect at least some of the features of M~theory, and
that bring exciting new geometric and particle physics ideas into
play. The RS2 models also provide a framework for exploring
holographic ideas that have emerged in M~theory. Most of the
recent progress on brane-world cosmology has been in RS-type
models. (Recent reviews are given in~\cite{m2,rev,lan}).

The field equations are~\cite{sms}
 \bea
^{(5)}\!G_{AB} &=& -\Lambda_\vd\, ^{(5)}\!g_{AB}\,,~(\mbox{bulk})\\
G_{\mu\nu} &=& - \Lambda g_{\mu\nu} + \kappa^2 T_{\mu\nu} +
6\frac{\kappa^2}{\sigma} {\cal S}_{\mu\nu} - {\cal E}_{\mu\nu}\,,
~\mbox{(brane)} \label{e:einstein1}
 \eea
where
 \bea
\kappa^2 &\equiv & \kappa^2_{4}={1\over6}\sigma\kappa^4_\vd \,,\\
\Lambda &\equiv& \Lambda_4 = {1\over 2}\left[
\Lambda_\vd+\kappa^2\sigma \right]\,,
 \eea
and the standard 4D conservation equations hold on the brane (when
there is only vacuum energy in the bulk):
 \be\label{lc}
\nabla^\nu T_{\mu\nu}=0\,.
 \ee

The induced field equations~(\ref{e:einstein1}) show two key
modifications to the standard 4D Einstein field equations arising
from extra-dimensional effects:
\begin{itemize}
\item ${\cal S}_{\mu\nu}\sim (T_{\mu\nu})^2$ is the high-energy
correction term, which is negligible for $\rho\ll\sigma$, but
dominant for $\rho\gg\sigma$:
 \be
{|\kappa^2{\cal S}_{\mu\nu}/\sigma |\over |\kappa^2
T_{\mu\nu}|}\sim {|T_{\mu\nu}|\over\sigma} \sim
{\rho\over\sigma}\,.
 \ee

\item ${\cal E}_{\mu\nu}$, the tracefree projection of the bulk
Weyl tensor on the brane, encodes corrections from 5D graviton
effects. These include the effects of the KK modes in the
linearized case, and the gravitational influence of the second
brane if there is one.

\end{itemize}

From the brane-observer viewpoint, the energy-momentum corrections
in ${\cal S}_{\mu\nu}$ are local, whereas the KK corrections in
${\cal E}_{\mu\nu}$ are nonlocal, since they incorporate 5D
gravity wave modes. These nonlocal corrections cannot be
determined purely from data on the brane. They are constrained by
the 4D contracted Bianchi identities ($\nabla^\nu G_{\mu\nu}=0$),
applied to Eq.~(\ref{e:einstein1}):
 \be \label{nlc}
\nabla^\mu{\cal
E}_{\mu\nu}={6\kappa^2\over\sigma}\,\nabla^\mu{\cal S}_{\mu\nu}\,.
 \ee
This shows qualitatively how 1+3 spacetime variations in the
matter-radiation on the brane can source KK modes. The 9
independent components in the tracefree ${\cal E}_{\mu\nu}$ are
reduced to 5 degrees of freedom by Eq.~(\ref{nlc}); these arise
from the 5 polarizations of the 5D graviton.

The trace free ${\cal E}_{\mu\nu}$ contributes an effective
``dark" radiative energy-momentum on the brane, with energy
density $\cu$, pressure $\cu/3$, momentum density $\cq_\mu $ and
anisotropic stress $\cp_{\mu\nu}$:
 \be
-{1\over\kappa^2} {\cal E}_{\mu\nu} = \cu\left({4\over3} u_\mu
u_\nu +{ {1\over3}} g_{\mu\nu}\right)+ {\cq_\mu } u_{\nu} +
{\cq_\nu } u_{\mu}+\cp_{\mu\nu}\,.
 \ee
We can think of this as a KK or Weyl ``fluid". The brane ``feels"
the bulk gravitational field through this effective fluid. The RS
models have a Minkowski brane in an AdS$_5$ bulk. This bulk is
also compatible with an FRW brane. However, the most general
vacuum bulk with a Friedmann brane is Schwarzschild-anti de Sitter
spacetime~\cite{birk}. It follows from the FRW symmetries that
$\cq_\mu =0=\cp_{\mu\nu}$, while $\cu=0$ only if the mass of the
black hole in the bulk is zero. The presence of the bulk black
hole generates via Coulomb effects the dark radiation on the
brane.

The brane-world corrections can conveniently be consolidated into
an effective total energy density, pressure, momentum density and
anisotropic stress. In the case of a perfect fluid (or minimally
coupled scalar field),
\begin{eqnarray}
\rho^{\text{tot}} &=& \rho\left(1 +\frac{\rho}{2\sigma} +
\frac{\cu}{\rho} \right)\,,\label{rtot} \\ \label{ptot}
p^{\text{tot }} &=& p  + \frac{\rho}{2\sigma}
(2p+\rho)+\frac{\cu}{3}\;, \\ q^{\text{tot }}_\mu  &=& \cq_\mu \;, \\
\label{e:pressure2} \pi^{\text{tot }}_{\mu\nu} &=& \cp_{\mu\nu}\;.
\end{eqnarray}

The (local) conservation equations~(\ref{lc}) are
\begin{eqnarray}
&&\dot{\rho}+\Theta(\rho+p)=0\,,\label{pc1}\\ && \nabla_i
p+(\rho+p)A_i =0\,,\label{pc2}
\end{eqnarray}
where $\Theta$ is the volume expansion rate ($=3H$ in FRW) and
$A_i$ is the 4-acceleration. The nonlocal conservation
equations~(\ref{nlc}) give~\cite{m1}
\begin{eqnarray}
&& \dot{\cu}+{{4\over3}}\Theta{\cu}+\nabla^i \cq_i =0\,,
\label{pc1'}\\&& \dot{q}^{\cal E}_i+{{4\over3}}\Theta\cq_i
+{{1\over3}}\nabla_i {\cu}+{{4\over3}}{\cu}A_i  +\nabla^j \cp_{ij}
=-{(\rho+p)\over\sigma} \nabla_i \rho\,,\label{pc2'}
\end{eqnarray}
where we have linearized about an FRW background.

\section{The background cosmology}

For an FRW brane, Eq.~(\ref{pc2'}) is trivially satisfied, while
Eq.~(\ref{pc1'}) gives the dark radiation solution
\begin{equation}\label{dr}
{\cu}=\rho_{{\cal E}\,0}\left({a_0\over a}\right)^4\,.
\end{equation}
In natural static coordinates, the Schwarzschild-AdS$_5$ metric
for an FRW brane-world is
 \bea
{}^\vu ds^2 &=& -F(R)dT^2+{dR^2 \over F(R)}+R^2\left({dr^2 \over
1- Kr^2}+r^2d\Omega^2\right)\,,\label{sads}\\ F(R) &=& K -{C \over
R^2}+ {R^2\over \ell^2}\,,\label{sads2}
 \eea
where $K=0,\pm1$ is the FRW curvature index and $C$ is the mass
parameter of the black hole at $R=0$ (note that the 5D
gravitational potential has $R^{-2}$ behaviour). The FRW brane
moves radially along the 5th dimension, with $R=a(T)$, where $a$
is the FRW scale factor, and the junction conditions determine the
velocity via the modified Friedmann equation~(\ref{mf}). We can
interpret the expansion of the universe as motion of the brane
through the static bulk.

The velocity of the brane is coordinate-dependent, and can be set
to zero. We can use Gaussian normal coordinates, in which the
brane is fixed but the bulk metric is not manifestly
static~\cite{bdel}:
 \be\label{gnm}
{}^\vu ds^2 = -N^2(t,y)dt^2+A^2(t,y)\left[{dr^2 \over 1-
Kr^2}+r^2d\Omega^2\right]+dy^2\,.
 \ee
Here $a(t)=A(t,0)$ is the scale factor on the FRW brane at $y=0$,
and $t$ may be chosen as proper time on the brane, so that
$N(t,0)=1$. In the case where there is no bulk black hole ($C=0$),
the metric functions are
 \bea
N &=& {\dot {A}(t,y) \over \dot {a}(t)}\,,\label{gnm1} \\ A &=&
a(t)\left[ \cosh \left({y \over \ell}\right) -
\left\{1+{\rho(t)\over\sigma}\right\}\sinh \left({|y| \over
\ell}\right)\right]\,.\label{gnm2}
 \eea
Again, the junction conditions determine the modified Friedmann
equation~\cite{bdel}
 \be\label{mf}
H^2 = \frac{\kappa^2}{3} \rho\left(1+{\rho\over 2\sigma}\right)
+{C\over a^4}+ \frac{1}{3} \Lambda - \frac{K}{a^2} \,,
 \ee
and by Eq.~(\ref{dr}),
 \be
C= \frac{\kappa^2}{3}\rho_{{\cal E}\,0}a_0^4\,.
 \ee
The Friedmann and matter energy conservation equations yield
 \be
\dot H= - {\kappa^2\over 2}(\rho+p)\left(1+ {\rho\over
\sigma}\right)-2{C\over a^4}+{K\over a^2}\,.
 \ee

The additional effective relativistic degree of freedom in dark
radiation is constrained by nucleosynthesis and CMB observations
to be no more than $\sim$5\% of the radiation energy
density~\cite{lmsw,dr}:
 \be
\left. {\cu \over \rho_{\rm rad}}\right|_{\rm nuc} \lesssim 0.05
 \ee
The other modification to the Hubble rate is via the high-energy
correction $\rho/\sigma$. In order to recover the observational
successes of general relativity, the high-energy regime where
significant deviations occur must take place before
nucleosynthesis, i.e., cosmological observations impose the lower
limit $\sigma > (1~{\rm MeV})^4$. This is much weaker than the
limit from table-top experiments:
 \be
\ell<0.1~{\rm mm}~\Rightarrow~ \sigma > (1~{\rm TeV})^4~{\rm and}~
M_\vd
> 10^5~{\rm TeV}\,.
 \ee

The background dynamics of brane-world cosmology are simple
because the FRW symmetries simplify the bulk and rule out nonlocal
effects. But perturbations on the brane in general release the
nonlocal KK modes. Then the 5D bulk perturbation equations must be
solved in order to solve for perturbations on the brane. These 5D
equations are partial differential equations for the 3-Fourier
modes, with complicated initial and boundary conditions.

The theory of gauge-invariant perturbations in brane-world
cosmology has been extensively investigated and developed (see
references given in the reviews~\cite{m2,rev,lan}) and is
qualitatively well understood. The key remaining task is
integration of the coupled brane-bulk perturbation equations with
appropriate initial/ boundary conditions. Special cases have been
solved, where these equations effectively decouple, as in the next
section, and approximation schemes have recently been
developed~\cite{sod,koy,rbbd,kkt,elmw} for the more general cases
where the coupled system must be solved. From the brane viewpoint,
the bulk effects, i.e., the high-energy corrections and the KK
modes, act as source terms for the brane perturbation equations.
At the same time, perturbations of matter on the brane can
generate KK modes (i.e., emit 5D gravitons into the bulk) which
propagate in the bulk and can subsequently interact with the
brane. This nonlocal interaction amongst the perturbations is at
the core of the complexity of the problem.

\section{Brane-world inflation}
%%%%%%%%%%%%%%%%%%%%%%%%%%%%%%%%%%%

In RS2-type brane-worlds, where the bulk has only a vacuum energy,
inflation on the brane must be driven by a 4D scalar field $\phi$
trapped on the brane~\cite{mwbh,inf}. (In more general
brane-worlds, where the bulk contains a 5D scalar field, it is
possible that the 5D field induces inflation on the brane via its
effective projection~\cite{hs}. More exotic possibilities arise
from the interaction between two branes, including possible
collision, which is mediated by a 5D scalar field and which can
induce either inflation~\cite{kss} or a hot big-bang radiation
era, as in the ``ekpyrotic" or cyclic scenario~\cite{ek}.)

High-energy brane-world modifications to the dynamics of inflation
provide increased Hubble damping, since $\rho\gg\sigma$ implies
$H$ is larger for a given energy than in 4D general
relativity~\cite{mwbh}. This makes slow-roll inflation possible
even for potentials that would be too steep in standard
cosmology~\cite{mwbh,steep,hulid1}.

The field satisfies the standard Klein-Gordon equation and the
modified Friedmann equation, with
$\rho={1\over2}\dot{\phi}^2+V(\phi)$ and
$p={1\over2}\dot{\phi}^2-V(\phi)$. The condition for inflation is
\begin{equation}
\label{endinf} \dot\phi^2 - V + \left[{{1\over2}\dot\phi^2 + V
\over \sigma}\left({5\over4}\dot\phi^2-{1\over2}V\right)\right] <
0 \,,
\end{equation}
which reduces to the general relativity result, $\dot{\phi}^2<V$,
when $\rho = {1\over2}\dot\phi^2+V \ll\sigma$. In the slow-roll
approximation,
\begin{eqnarray}
H^2 &\approx&  {\kappa^2\over3} V\left[ 1+{V\over2\sigma}
\right]\,,
 \label{7}\\
\dot\phi &\approx & -{V'\over 3H}\,. \label{8}
\end{eqnarray}
The brane-world correction term $V/\sigma$ in Eq.~(\ref{7}) serves
to enhance the Hubble rate for a given potential energy, relative
to general relativity. Thus there is enhanced Hubble `friction' in
Eq.~(\ref{8}), and brane-world effects will reinforce slow-roll at
the same potential energy. We can see this by defining slow-roll
parameters that reduce to the standard parameters in the
low-energy limit:
\begin{eqnarray}
\label{epsilon} \epsilon &\equiv& -{\dot H \over H^2}={M_{\rm p}^2
\over 16\pi} \left( {V' \over V} \right)^2 \left[ {1+V/\sigma
\over(1+V/2\sigma)^2} \right] \,,\label{10}\\ \label{eta} \eta
&\equiv & -{\ddot\phi \over H \dot\phi}={M_{\rm p}^2 \over 8\pi}
\left( {V'' \over V} \right) \left[ {1 \over 1+V/2\sigma} \right]
\,.\label{11}
\end{eqnarray}
Self-consistency of the slow-roll approximation then requires
$\epsilon,|\eta|\ll 1$. At low energies, $V\ll\sigma$, the
slow-roll parameters reduce to the standard form. However at high
energies, $V\gg\sigma$, the extra contribution to the Hubble
expansion helps damp the rolling of the scalar field and the new
factors in square brackets become $\approx\sigma/V$:
 \bea
\epsilon\approx\epsilon_{\rm gr}\left[{ {4\sigma\over
V}}\right]\,,~ \eta\approx\eta_{\rm gr}\left[{ {2\sigma\over
V}}\right],
 \eea
where $\epsilon_{\rm gr},\eta_{\rm gr}$ are the standard general
relativity slow-roll parameters. In particular, this means that
steep potentials which do not give inflation in general
relativity, can inflate the brane-world at high energy and then
naturally stop inflating when $V$ drops below $\sigma$. These
models can be constrained because they typically end inflation in
a kinetic-dominated regime and thus generate a blue spectrum of
gravitational waves, which can disturb
nucleosynthesis~\cite{steep}. They also allow for the novel
possibility that the inflaton could act as dark matter or
quintessence at low energies~\cite{steep,dq}.

The key test of any modified gravity theory during inflation, will
be the spectrum of perturbations produced due to quantum
fluctuations of the fields about their homogeneous background
values. In general, perturbations on the brane are coupled to bulk
metric perturbations, and the problem is very complicated. However
on large scales on the brane, the density and curvature
perturbations decouple from the bulk metric
perturbations~\cite{m1,mwbh,lmsw} (see the next section). Thus we
are justified in neglecting the bulk metric perturbations when
computing the density perturbations.

To quantify the amplitude of scalar (density) perturbations we
evaluate the usual gauge-invariant quantity
\begin{equation}
\label{defzeta} \zeta \equiv {\cal R}-{H\over\dot\rho}\delta\rho
\,,
\end{equation}
which reduces to the curvature perturbation, ${\cal R}$, on
uniform density hypersurfaces ($\delta\rho=0$). This is conserved
on large scales for purely adiabatic perturbations, as a
consequence of energy conservation (independently of the field
equations)~\cite{wmll}. The curvature perturbation on uniform
density hypersurfaces is given in terms of the scalar field
fluctuations on spatially flat hypersurfaces, $\delta\phi$, by
\begin{equation}
\zeta = H\,{\delta\phi\over\dot\phi} \,. \label{9}
\end{equation}
The field fluctuations at Hubble crossing ($k=aH$) in the
slow-roll limit are given by
$\langle\delta\phi^2\rangle\approx\left({H/2\pi} \right)^2$, a
result for a massless field in de Sitter space that is also
independent of the gravity theory~\cite{wmll}. For a single scalar
field the perturbations are adiabatic and hence the curvature
perturbation $\zeta$ can be related to the density perturbations
when modes re-enter the Hubble scale during the matter dominated
era, which is given by $A_{\rm s}^2 = 4\langle \zeta^2
\rangle/25$. Using the slow-roll equations and Eq.~(\ref{9}), this
gives
\begin{equation}
\label{AS} A_{\rm s}^2 \approx \left . \left({512\pi\over75 M_{\rm
p} ^6}\, {V^3 \over V^{\prime2}}\right)\left[ {2\sigma + V \over
2\sigma} \right]^3 \right|_{k=aH}\,.
\end{equation}
Thus the amplitude of scalar perturbations is {\em increased}
relative to the standard result at a fixed value of $\phi$ for a
given potential.

The scale-dependence of the perturbations is described by the
spectral tilt
\begin{equation}
n_{\rm s}-1\equiv {d\ln A_{\rm s}^2 \over d\ln k} \approx
-6\epsilon + 2\eta \,,\label{15}
\end{equation}
where the slow-roll parameters are given in Eqs.~(\ref{epsilon})
and~(\ref{eta}). Because these slow-roll parameters are both
suppressed by an extra factor $\sigma/V$ at high energies, we see
that the spectral index is driven towards the Harrison-Zel'dovich
spectrum, $n_{\rm s}\to1$, as $V/\sigma\to\infty$; however, this
does not necessarily mean that the brane-world case is closer to
scale-invariance than the general relativity case. In comparing
the high-energy brane-world case to the standard 4D case, we
implicitly require the same potential energy. However, precisely
because of the high-energy effects, large-scale perturbations will
be generated at different values of $V$ than in the standard case,
specifically at lower values of $V$, closer to the reheating
minimum. Thus there are two competing effects, and it turns out
that the shape of the potential determines which is the dominant
effect~\cite{lidsmi}.

\begin{figure}[!bth]\label{gw}
\begin{center}
\includegraphics[height=4in, width=4in]{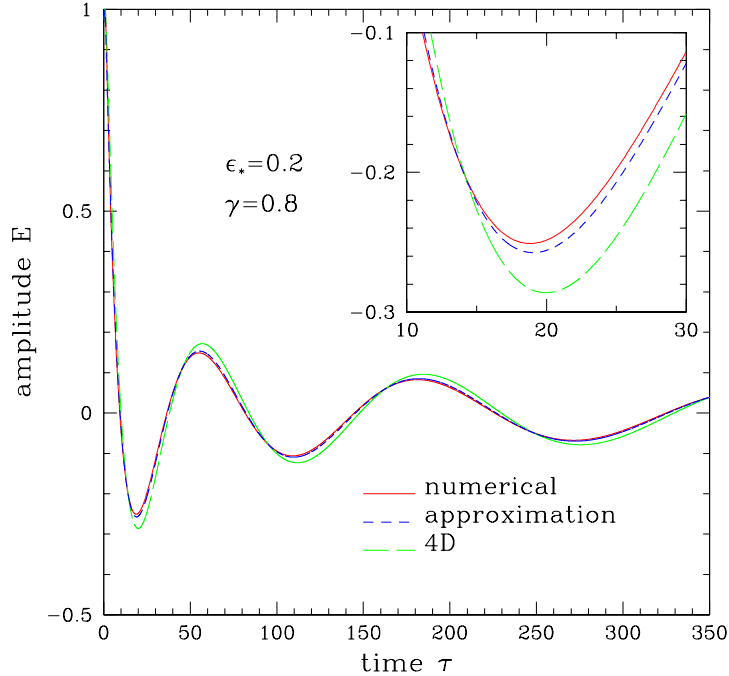}
\caption{The damping of cosmological gravity waves on horizon
re-entry due to massive mode generation. The solid curve is the
numerical solution, the short-dashed curve the low-energy
approximation, and the long-dashed curve the standard general
relativity solution. $\epsilon_*=\rho_0/\sigma$ and $\gamma$ is a
parameter giving the location of the regulator brane.
(From~\cite{kkt}.) }
\end{center}
\end{figure}

High-energy inflation on the brane also generates a zero-mode (4D
graviton mode) of tensor perturbations, and stretches it to
super-Hubble scales. This zero-mode has the same qualitative
features as in general relativity, remaining frozen at constant
amplitude while beyond the Hubble horizon. Its amplitude is
enhanced at high energies, although the enhancement is much less
than for scalar perturbations~\cite{lmw}:
 \bea
A_{\rm t}^2 &\approx& \left({32V\over 75 M_{\rm p}^2}\right)
\left[ {3V^2 \over 4\sigma^2}\right],\label{higw}\\  {A_{\rm
t}^2\over A_{\rm s}^2} &\approx& \left({M_{\rm p}^2\over
16\pi}\,{V'^2\over V^2}\right) \left[ {6\sigma\over
V}\right].\label{ten}
 \eea
Equation~(\ref{ten}) means that brane-world effects suppress the
large-scale tensor contribution to CMB anisotropies. The tensor
spectral index at high energy has a smaller magnitude than in
general relativity,
 \be
n_{\rm t}=-3\epsilon\,,
 \ee
but remarkably the same consistency relation as in general
relativity holds~\cite{hulid1}:
 \be
n_{\rm t} = -2{A_{\rm t}^2\over A_{\rm s}^2}\,.
 \ee

The massive KK modes of tensor perturbations of a de Sitter brane
have a mass gap~\cite{gs,lmw,grs,fk}: $m>3H/2$. These massive
modes remain in the vacuum state during slow-roll
inflation~\cite{lmw,grs}. The evolution of the super-Hubble zero
mode is the same as in general relativity, so that high-energy
brane-world effects in the early universe serve only to rescale
the amplitude. However, when the zero mode re-enters the Hubble
horizon, massive KK modes can be excited, leading to a loss of
energy from the zero mode, which can be estimated at low
energies~\cite{kkt,elmw} (see Fig.~2).

Vector perturbations in the bulk metric can support vector metric
perturbations on the brane, even in the absence of matter
perturbations~\cite{m1}. However, there is no normalizable zero
mode, and the massive KK modes stay in the vacuum state during
brane-world inflation~\cite{bmwv}. Therefore, as in general
relativity, we can neglect vector perturbations in inflationary
cosmology.

\section{Curvature perturbations on large scales}

The curvature perturbation ${\cal R}$ on uniform density surfaces
is associated with the gauge-invariant quantity in
Eq.~(\ref{defzeta}). This is defined for matter on the brane, in
the usual way. Similarly, for the Weyl ``fluid" if $\cu\neq0$ in
the background, the curvature perturbation on hypersurfaces of
uniform dark energy density is
 \be
\zeta_{\cal E}={\cal R}+ {\delta\cu \over 4\cu}\,.
 \ee
On large scales, the dark energy conservation
equation~(\ref{pc1'}) implies
 \be
(\delta\cu)^{\displaystyle{\cdot}}+4H\delta\cu+ 4\cu \dot{\cal
R}=0\,,
 \ee
which leads to
 \be
\dot{\zeta}_{\cal E}=0\,.
 \ee
For adiabatic matter perturbations, by the perturbed matter energy
conservation equation,
 \be
(\delta\rho)^{\displaystyle{\cdot}}+3H(\delta\rho+\delta p)+
3(\rho +p) \dot{\cal R}=0\,,
 \ee
we find
 \be
\dot\zeta=0\,.
 \ee
This is independent of brane-world modifications to the field
equations, since it depends on energy conservation only. For the
total, effective fluid, the curvature perturbation is defined as
follows~\cite{lmsw}: if $\cu\neq0$ in the background,
 \bea
{\zeta}^{\rm tot} &=& {\zeta}+\left[{4\cu \over 3(\rho+ p)(1+
\rho/\sigma)+4\cu}\right] \left(\zeta_{\cal E}-\zeta\right)\,,
 \eea
and if $\cu=0$ in the background,
 \bea
{\zeta}^{\rm \,tot} &=& {\zeta}+ {\delta\cu \over
3(\rho+p)(1+\rho/\sigma)}\,, \\ \delta\cu &=& {\delta C \over a^4}
\,,
 \eea
where $\delta C$ is constant. It follows that the curvature
perturbations on large scales can be found on the brane without
solving for the bulk metric perturbations.

Although the density and curvature perturbations can be found on
super-Hubble scales, the Sachs-Wolfe effect requires
$\cp_{\mu\nu}$ in order to translate from density/ curvature to
metric perturbations. In the 4D longitudinal gauge of the metric
perturbation formalism, the gauge-invariant curvature and metric
perturbations on large scales are related by~\cite{lmsw}
 \bea
\zeta^{\rm tot} &=& {\cal R}-{H \over \dot H}\left( {\dot{\cal R}
\over H}-\psi\right) \,, \label{curv}
\\ {\cal R}+\psi &=& -\kappa^2a^2\delta \pi_{\cal E}\,,\label{metcurv}
 \eea
where the radiation anisotropic stress on large scales is
neglected, as in general relativity, and $\delta\pi_{\cal E}$ is
the scalar potential for $\cp_{\mu\nu}$. In 4D general relativity,
the right hand side of Eq.~(\ref{metcurv}) is zero. The
(non-integrated) Sachs-Wolfe formula has the same form as in
general relativity:
 \be \label{swd}
{\delta T\over T}\Big|_{\rm now}=(\zeta_{\rm rad}+\psi-{\cal
R})|_{\rm dec}\,.
 \ee
The brane-world corrections to the general relativistic
Sachs-Wolfe effect are then given by~\cite{lmsw}
 \be\label{sachsw}
{\delta T\over T} = \left({\delta T\over T}\right)_{\rm gr}
-{8\over 3}\left({\rho_{\rm rad}\over \rho_{\rm
cdm}}\right)S_{\cal E}-\kappa^2a^2\delta \pi_{\cal E}
+{2\kappa^2\over a^{5/2}}\int da\,\, a^{7/2}\,\delta\pi_{\cal E}
\,,
 \ee
where $S_{\cal E}$ is the KK entropy perturbation (determined by
$\delta\cu$). The KK term $\delta\pi_{\cal E}$ cannot be
determined by the 4D brane equations, so that $\delta T/T$ cannot
be evaluated on large scales without solving the 5D equations.
Equation~(\ref{sachsw}) has been generalized to the 2-brane case,
in which the radion makes a contribution to the Sachs-Wolfe
effect~\cite{ksw}.

The presence of the KK (Weyl, dark) component has essentially two
possible effects.
\begin{itemize}
\item A contribution from the KK entropy perturbation $S_{\cal E}$ that
is similar to an extra isocurvature contribution. \item A
contribution from the KK anisotropic stress $\delta\pi_{\cal E}$.
In the absence of anisotropic stresses, the curvature perturbation
$\zeta^{\rm tot}$ would be sufficient to determine the metric
perturbation ${\cal R}$ and hence the large-angle CMB
anisotropies, via Eqs.~(\ref{curv}), (\ref{metcurv}) and
(\ref{swd}). However bulk gravitons can also generate anisotropic
stresses which, although they do not affect the large-scale
curvature perturbation $\zeta^{\rm tot}$, can affect the relation
between $\zeta^{\rm tot}$, ${\cal R}$ and $\psi$, and hence can
affect the CMB anisotropies at large angles.
\end{itemize}

\section{Brane-world CMB anisotropies}
%%%%%%%%%%%%%%%%%%%%%%%%%%%%%%%%%%%%%%%%%%%%%

Recently, the anisotropies in the CMB for RS-type brane-world
cosmologies have been calculated using a low-energy
approximation~\cite{koy}. The basic idea of the low-energy
approximation~\cite{sod} is to use a gradient expansion to exploit
the fact that, during most of the history of the universe, the
curvature scale on the observable brane is much greater than the
curvature scale of the bulk ($\ell<1~$mm):
 \bea
&& L\sim |R_{\mu\nu\alpha\beta}|^{-1/2} \gg \ell \sim |\,^\vu
R_{ABCD}|^{-1/2} \nonumber\\ && \Rightarrow |\nabla_\mu|\sim
L^{-1} \ll |\partial_y|\sim \ell^{-1}\,. \label{grad}
 \eea
These conditions are equivalent to the low energy regime, since
$\ell^2\propto \sigma^{-1}$ and $|R_{\mu\nu\alpha\beta}|\sim
|T_{\mu\nu}|$:
 \be
{\ell^2 \over L^2} \sim {\rho \over \sigma} \ll 1\,.
 \ee
Using Eq.~(\ref{grad}) to neglect appropriate gradient terms in an
expansion in $\ell^2/L^2$, the low-energy equation
 \be
\nabla^\nu {\cal E}_{\mu\nu}=0\,,
 \ee
can be solved. However, two boundary conditions are needed to
determine all functions of integration. This is achieved by
introducing a second brane, as in the RS1 scenario. This brane is
to be thought of either as a regulator brane, whose backreaction
on the observable brane is neglected (which will only be true for
a limited time), or as a shadow brane with physical fields, which
have a gravitational effect on the observable brane.

The background is given by low-energy FRW branes with tensions
$\pm\sigma$, proper times $t_\pm$, scale factors $a_\pm$, energy
densities $\rho_\pm$ and pressures $p_\pm$, and dark radiation
densities $\rho_{{\cal E}\,\pm}$. The physical distance between
the branes is $\ell \bar{d}(t)$, and
 \be
{d\over dt_-}=e^{\bar{d}}\,{d \over dt_+}\,,~
a_-=a_+e^{-{\bar{d}}}\,,~ H_-=e^{\bar{d}}\left(H_+- \dot
{\bar{d}}\right)\,,~ \rho_{{\cal E}\,-}=e^{4{\bar{d}}}\rho_{{\cal
E}\,+}\,.
 \ee
Then the background dynamics are given by
 \bea
&& H_\pm^2 = \pm{\kappa^2 \over 3} \left(\rho_\pm \pm\rho_{{\cal
E}\,\pm} \right)\,,
\\
&& \ddot {\bar{d}}+3H_+\dot {\bar{d}}-\dot{{\bar{d}}}^2 =
{\kappa^2 \over 6}\left[ \rho_+-3p_+ +e^{2{\bar{d}}}(\rho_--3p_-)
\right]\,.\label{dbar}
 \eea
The dark energy obeys $\rho_{{\cal E}\,+} =C/a_+^4$, where $C$ is
a constant. From now on, we drop the +-subscripts which refer to
the physical, observed quantities.

The perturbed metric on the observable (positive tension) brane is
described, in longitudinal gauge, by the metric perturbations
$\psi$ and ${\cal R}$, and the perturbed radion is $d= {\bar{d}}+
N$. The approximation for the KK (Weyl) energy-momentum tensor on
the observable brane is~\cite{koy}
 \bea
{\cal E}^{\mu}{}_{\nu}& =& \frac{2}{e^{2d}-1} \left[
-\frac{\kappa^2}{2}\left( T^{\mu}_{\nu}+e^{-2 d}T^{\mu}_{-~
\nu}\right) \right.\nonumber \\ &&\left.~~{} -
\nabla^{\mu}\nabla_{\nu} d + \delta^{\mu}_{\nu} \nabla^2 d
-\left\{ \nabla^{\mu}d \nabla_{\nu} d + \frac{1}{2}
\delta^{\mu}_{\nu} (\nabla d)^2 \right\} \right], \label{solE}
 \eea
and the field equations on the observable brane can be written in
scalar-tensor form as
\begin{eqnarray}
G^\mu{}_\nu & = & \frac{\kappa^2}{\chi}
T^\mu_{\nu}+\frac{\kappa^2(1-\chi)^2}{\chi} T^\mu_{-~\nu}
\nonumber \\
& &~{} +\frac{1}{\chi}\left(\nabla^\mu \nabla_\nu
\chi-\delta^\mu_\nu \nabla^2\chi \right)
+\frac{\omega(\chi)}{\chi^2}\left[\nabla^\mu \chi \nabla_\nu
\chi-\frac{1}{2}\delta^\mu_\nu (\nabla \chi)^2 \right],
\end{eqnarray}
where
 \be
\chi=1-e^{-2d}\,,~\omega(\chi)=\frac{3}{2}\frac{\chi}{1-\chi}\,.
 \ee

\begin{figure}[bth]\label{koyfig}
\begin{center}
\includegraphics[height=4in, width=5in]{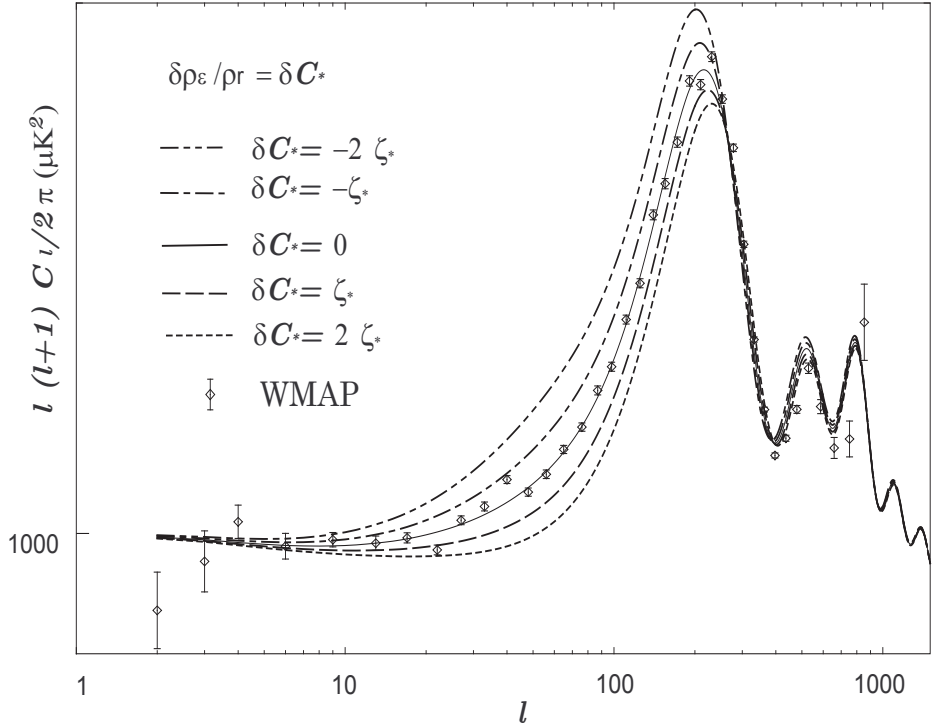}
\caption{The CMB power spectrum with brane-world effects, for
different values of the parameter $\delta C_*$ (which is the
large-scale dark radiation fluctuation $\delta\cu$ as a proportion
of the large-scale curvature perturbation for matter $\zeta_*$).
(From~\cite{koy}.) }
\end{center}
\end{figure}

The perturbation equations can then be derived as generalizations
of the standard equations. The trace part of the perturbed field
equation shows that the radion perturbation determines the crucial
quantity, $\delta\pi_{\cal E}$:
 \be\label{rplusp}
{\cal R}+\psi = -{2 \over e^{2\bar d}-1}N=-\kappa^2a^2\delta
\pi_{\cal E}\,,
 \ee
where the last equality follows from Eq.~(\ref{metcurv}). A new
set of variables $\varphi_\pm, E$ turns out be very
useful~\cite{ksw,koy}:
\begin{eqnarray}
{\cal R} &=&  -\varphi_+ - {a^{2}\over k^2} H \dot{E} +
\frac{1}{3} E\,,
\nonumber\\
\psi &=& - \varphi_+ - {a^{2}\over k^2} (\ddot{E}+ 2H \dot{E})\,,
\nonumber\\
N &=& \varphi_- - \varphi_+ -  {a^{2}\over k^2}\dot{\bar d}
\dot{E}\,.
\end{eqnarray}
The variable $E$ determines the metric shear anisotropy in the
bulk, whereas $\varphi_\pm$ give the brane displacements, in
transverse traceless gauge. The latter variables have a simple
relation to the curvature perturbations on large
scales~\cite{ksw,koy} (restoring the +-subscripts):
 \be
\zeta^{\rm tot}_\pm =-\varphi_\pm +{H_\pm^2 \over \dot H_\pm}
\left({\dot{\varphi}_\pm \over H_\pm}+ \varphi_\pm \right)\,,
 \ee
where $\dot{f}_\pm\equiv df_\pm/dt_\pm$.

The simplest model has
 \be
\cu=0=\dot{\bar d}
 \ee
in the background, with $p_-/\rho_-=p/\rho$. By Eq.~(\ref{dbar}),
it follows that
 \be
\rho_-=-\rho e^{2\bar d}\,,
 \ee
i.e., the matter on the regulator brane must have fine-tuned and
negative energy density to prevent the regulator brane from moving
in the background. The regulator brane is assumed to be very far
from the physical brane, so that we can neglect its effects over a
cosmological time-scale. With these assumptions, and further
assuming adiabatic perturbations for the matter, there is only one
independent brane-world parameter, i.e., the parameter measuring
dark radiation fluctuations:
 \be
\delta C_*= {\delta\cu \over \rho_{\rm rad}}\,.
 \ee

This has a remarkable consequence on large scales: the Weyl
anisotropic stress $\delta\pi_{\cal E}$ terms in the Sachs-Wolfe
formula Eq.~(\ref{sachsw}) cancel the entropy perturbation from
dark radiation fluctuations, so that there is no difference on the
largest scales from the standard general relativity power
spectrum. On small scales, beyond the first acoustic peak, the
brane-world corrections are negligible. On scales up to the first
acoustic peak, brane-world effects can be significant, changing
the height and the location of the first peak. These features are
apparent in Fig.~3. However, it is not clear to what extent these
features are general brane-world features (within the low-energy
approximation), and to what extent they are consequences of the
simple assumptions imposed on the background. Further work remains
to be done. (A related low-energy approximation, using the moduli
space approximation, has been developed for certain 2-brane models
with bulk scalar field~\cite{rbbd}.)

\section{Conclusion}
%%%%%%%%%%%%%%%%%%%%

Simple brane-world cosmologies of RS type provide a rich
phenomenology for exploring some of the ideas that are emerging
from M~theory. At the same time, brane-world gravity opens up
exciting prospects for subjecting M~theory ideas to the
increasingly stringent tests provided by high-precision
astronomical observations. Recent progress in tackling the massive
KK modes for tensor and scalar perturbations, and in particular
the development of an approximation scheme for computing CMB
anisotropies, mean that these goals are brought closer to
realization. On this basis, the perturbation analysis can be
extended to cover more realistic brane-world models, and
ultimately M~theory models.

\[ \]
{\bf Acknowledgments}

I thank the organisers for a very enjoyable and stimulating
symposium. I am supported by PPARC.


\begin{thebibliography}{0}

\bibitem{mtheory}
For recent reviews, see, e.g., J.H. Schwarz, astro-ph/0304507; J.
Polchinski, hep-th/0209105; R. Kallosh, hep-th/0205315.

\bibitem{add}
N. Arkani-Hamed, S. Dimopoulos, G. Dvali, Phys. Lett. {\bf B429},
263 (1998) [hep-ph/9803315]; I. Antoniadis, N. Arkani-Hamed, S.
Dimopoulos, G. Dvali, Phys. Lett. {\bf B436}, 257 (1998)
[hep-ph/9804398].

\bibitem{cav}
M. Cavaglia, Int. J. Mod. Phys. {\bf A18}, 1843 (2003)
[hep-ph/0210296].

\bibitem{hv}
P. Horava, E. Witten, Nucl. Phys. {\bf B460}, 506 (1996)
[hep-th/9510209].

\bibitem{low}
A. Lukas, B.A. Ovrut, K.S. Stelle, D. Waldram, Phys. Rev. D{\bf
59}, 086001 (1999) [hep-th/9803235]; A. Lukas, B.A. Ovrut, D.
Waldram, Phys. Rev. D{\bf 60}, 086001 (1999) [hep-th/9806022];
ibid., {\bf 61}, 023506 (2000) [hep-th/9902071].

\bibitem{rs1}
L. Randall, R. Sundrum, Phys. Rev. Lett. {\bf 83}, 3370 (1999)
[hep-ph/9905221].

\bibitem{rs2}
L. Randall, R. Sundrum, Phys. Rev. Lett. {\bf 83}, 4690 (1999)
[hep-th/9906064].

\bibitem{m2}
R. Maartens, gr-qc/0312059.

\bibitem{rev}
P. Brax, C. van de Bruck, Class. Quantum Grav. {\bf 20}, R201
(2003) [hep-th/0303095].

\bibitem{lan}
D. Langlois, astro-ph/0301021.

\bibitem{sms} T. Shiromizu, K. Maeda, M. Sasaki,
Phys. Rev. D{\bf 62}, 024012 (2000) [gr-qc/9910076].

\bibitem{birk}
S. Mukohyama, T. Shiromizu, K. Maeda, Phys. Rev. D{\bf 62}, 024028
(2000) [hep-th/9912287]; P. Bowcock, C. Charmousis, R. Gregory,
Class. Quantum Grav. {\bf 17}, 4745 (2000) [hep-th/0007177].

\bibitem{m1}
R. Maartens, Phys. Rev. D{\bf 62}, 084023 (2000) [hep-th/0004166].

\bibitem{bdel}
P. Binetruy, C. Deffayet, U. Ellwanger, D. Langlois, Phys. Lett.
{\bf B477}, 285 (2000) [hep-th/9910219].

\bibitem{lmsw}
D. Langlois, R. Maartens, M. Sasaki, D. Wands, Phys. Rev. D{\bf
63}, 084009 (2001) [hep-th/0012044].

\bibitem{dr}
J.D. Barrow, R. Maartens, Phys. Lett. {\bf B532}, 153 (2002)
[gr-qc/0108073]; K. Ichiki, M. Yahiro, T. Kajino, M. Orito, G.J.
Mathews, Phys. Rev. D{\bf 66}, 043521 (2002) [astro-ph/0203272];
J.D. Bratt, A.C. Gault, R.J. Scherrer, T.P. Walker, Phys. Lett.
{\bf B546}, 19 (2002) [astro-ph/0208133].

\bibitem{sod}
J. Soda, S. Kanno, Phys. Rev. D{\bf 66}, 083506 (2002)
[hep-th/0207029]; T. Wiseman, Class. Quantum Grav. {\bf 19}, 3083
(2002) [hep-th/0201127]; T. Shiromizu, K. Koyama, Phys. Rev. D{\bf
67}, 084022 (2003) [hep-th/0210066]; J. Soda, S. Kanno, Astrophys.
Space Sci. {\bf 283}, 639 (2003) [gr-qc/0209086].

\bibitem{koy}
K. Koyama, Phys. Rev. Lett. {\bf 91}, 221301 (2003)
[astro-ph/0303108].

\bibitem{rbbd}
C.S. Rhodes, C. van de Bruck, Ph. Brax, A.C. Davis,  Phys. Rev.
D{\bf 68}, 083511 (2003) [astro-ph/0306343]; P. Brax, C. van de
Bruck, A.-C. Davis, C.S. Rhodes, hep-ph/0309181.

\bibitem{kkt}
T. Hiramatsu, K. Koyama, A. Taruya, Phys. Lett. {\bf B578}, 269
(2004) [ hep-th/0308072].

\bibitem{elmw}
R. Easther, D. Langlois, R. Maartens, D. Wands, JCAP {\bf 10}, 014
(2003) [hep-th/0308078].

\bibitem{mwbh}
R. Maartens, D. Wands, B.A. Bassett, I.P.C. Heard, Phys. Rev.
D{\bf 62}, 041301 (2000) [hep-ph/9912464].

\bibitem{inf}
N. Kaloper, Phys. Rev. D{\bf 60}, 123506 (1999) [hep-th/9905210];
J.M. Cline, C. Grojean, G. Servant, Phys. Rev. Lett. {\bf 83},
4245 (1999) [hep-ph/9906523]; H. Stoica, S.-H. Henry Tye, I.
Wasserman, Phys. Lett. {\bf B482}, 205 (2000) [hep-th/0004126]; L.
Mendes, A.R. Liddle, Phys. Rev. D{\bf 62}, 103511 (2000)
[astro-ph/0006020]; A. Mazumdar, Phys. Rev. D{\bf 64}, 027304
(2001) [hep-ph/0007269]; S.C. Davis, W.B. Perkins, A.-C. Davis,
I.R. Vernon, Phys. Rev. D{\bf 63}, 083518 (2001) [hep-ph/0012223];
A.R. Liddle, A.N. Taylor, Phys. Rev. D{\bf 65}, 041301 (2002)
[astro-ph/0109412]; M.C. Bento, O. Bertolami, Phys. Rev. D{\bf
65}, 063513 (2002) [astro-ph/0111273]; M.C. Bento, O. Bertolami,
A.A. Sen, Phys. Rev. D{\bf 67}, 023504 (2003) [gr-qc/0204046];
ibid., 063511 (2003) [hep-th/0208124]; S. Mizuno, K. Maeda, K.
Yamamoto, Phys. Rev. D{\bf 67}, 024033 (2003) [hep-ph/0205292]; R.
Hawkins, J.E. Lidsey, Phys. Rev. D{\bf 68}, 083505 (2003)
[astro-ph/0306311]; K.E. Kunze, hep-th/0310200.

\bibitem{hs}
S. Kobayashi, K. Koyama, J. Soda, Phys. Lett. {\bf B501}, 157
(2001) [hep-th/0009160]; Y. Himemoto, M. Sasaki, Phys. Rev. D{\bf
63}, 044015 (2001) [gr-qc/0010035]; E.E. Flanagan, S.-H. Henry
Tye, I. Wasserman, Phys. Lett. {\bf B522}, 155 (2001)
[hep-th/0110070]; N. Sago, Y. Himemoto, M. Sasaki, Phys. Rev.
D{\bf 65}, 024014 (2002) [gr-qc/0104033]; Y. Himemoto, T. Tanaka,
M. Sasaki, Phys. Rev. D{\bf 65}, 104020 (2002) [gr-qc/0112027]; Y.
Himemoto, T. Tanaka, Phys. Rev. D{\bf 67}, 084014 (2003)
[gr-qc/0212114]; T. Tanaka, Y. Himemoto, Phys. Rev. D{\bf 67},
104007 (2003) [gr-qc/0301010]; B. Wang, L-H. Xue, X. Zhang, W-Y.P.
Hwang, Phys. Rev. D{\bf 67}, 123519 (2003) [hep-th/0301072]; K.
Koyama, K. Takahashi, Phys. Rev. D{\bf 67}, 103503 (2003)
[hep-th/0301165]; D. Langlois, M. Sasaki, Phys. Rev. D{\bf 68},
064012 (2003) [hep-th/0302069]; Y. Himemoto, M. Sasaki, Prog.
Theor. Phys. Suppl. {\bf 148}, 235 (2002) [gr-qc/0302054]; R.H.
Brandenberger, G. Geshnizjani, S. Watson,  Phys. Rev. D{\bf 67},
123510 (2003) [hep-th/0302222]; M. Minamitsuji, Y. Himemoto, M.
Sasaki, Phys. Rev. D{\bf 68}, 024016 (2003) [gr-qc/0303108]; S.
Kanno, J. Soda, hep-th/0303203; J. Martin, G.N. Felder, A.V.
Frolov, M. Peloso, L. Kofman, hep-th/0309001; A.V. Frolov, L.
Kofman, hep-th/0309002; P.R. Ashcroft, C. van de Bruck, A.-C.
Davis, astro-ph/0310643.

\bibitem{kss}
G. Dvali, S.-H.H. Tye, Phys. Lett. {\bf B450}, 72 (1999)
[hep-ph/9812483]; S. Kanno, M. Sasaki, J. Soda, Prog. Theor. Phys.
{\bf 109}, 357 (2003) [hep-th/0210250].

\bibitem{ek}
J. Khoury, B.A. Ovrut, P.J. Steinhardt, N. Turok, Phys. Rev. D{\bf
64}, 123522 (2001) [hep-th/0103239]; R. Kallosh, L. Kofman, A.
Linde, Phys. Rev. D{\bf 64}, 123523 (2001) [hep-th/0104073]; A.
Neronov, JHEP {\bf 11}, 007 (2001) [hep-th/0109090]; P.J.
Steinhardt, N. Turok, Phys. Rev. D{\bf 65}, 126003 (2002)
[hep-th/0111098]; D. Langlois, K. Maeda, D. Wands, Phys. Rev.
Lett. {\bf 88}, 181301 (2002) [gr-qc/0111013]; N.E. Mavromatos,
hep-th/0210008; A.J. Tolley, N. Turok, P.J. Steinhardt,
hep-th/0306109

\bibitem{steep}
E.J. Copeland, A.R. Liddle, J.E. Lidsey, Phys. Rev. D{\bf 64},
023509 (2001) [astro-ph/0006421];  A. S. Majumdar, Phys. Rev.
D{\bf 64}, 083503 (2001) [astro-ph/0105518]; V. Sahni, M. Sami, T.
Souradeep, Phys. Rev. D{\bf 65}, 023518 (2002) [gr-qc/0105121];
N.J. Nunes, E.J. Copeland, Phys. Rev. D{\bf 66}, 043524 (2002)
[astro-ph/0204115]; A.R. Liddle, L.A. Urena-Lopez, Phys. Rev.
D{\bf 68}, 043517 (2003) [astro-ph/0302054].

\bibitem{hulid1}
G. Huey, J.E. Lidsey, Phys. Lett. {\bf B514}, 217 (2001)
[astro-ph/0104006].

\bibitem{dq}
A. Albrecht, C.P. Burgess, F. Ravndal, C. Skordis, Phys. Rev.
D{\bf 65}, 123507 (2002) [astro-ph/0107573]; S. Mizuno, K. Maeda,
Phys. Rev. D{\bf 64}, 123521 (2001) [hep-ph/0108012]; J.E. Lidsey,
T. Matos, L.A. Urena-Lopez, Phys. Rev. D{\bf 66}, 023514 (2002)
[astro-ph/0111292]; C.P. Burgess, astro-ph/0207174; D. Seery, B.A.
Bassett, astro-ph/0310208.

\bibitem{wmll}
D. Wands, K. A. Malik, D. H. Lyth, A. R. Liddle, Phys. Rev. D{\bf
62},  043527 (2000) [astro-ph/0003278].

\bibitem{lidsmi}
A.R. Liddle, A.J. Smith, Phys. Rev. D{\bf 68}, 061301 (2003)
[astro-ph/0307017].

\bibitem{lmw}
D. Langlois, R. Maartens, D. Wands, Phys. Lett. {\bf B489}, 259
(2000) [hep-th/0006007].

\bibitem{gs}
J. Garriga, M. Sasaki, Phys. Rev. D{\bf 62}, 043523 (2000)
[hep-th/9912118].

\bibitem{grs}
D.S. Gorbunov, V.A. Rubakov, S.M. Sibiryakov, JHEP {\bf 10}, 15
(2001) [hep-th/0108017].

\bibitem{fk}
A. Frolov, L. Kofman, hep-th/0209133.

\bibitem{bmwv}
H.A. Bridgman, K.A. Malik, D. Wands, Phys. Rev. D{\bf 63}, 084012
(2001) [hep-th/0010133].

\bibitem{ksw}
K. Koyama, Phys. Rev. D{\bf 66}, 084003 (2002) [gr-qc/0204047].

\bibitem{bmw}
H.A. Bridgman, K.A. Malik, D. Wands, Phys. Rev. D{\bf 65}, 043502
(2002) [astro-ph/0107245].

\bibitem{gwtan}
T. Kobayashi, H. Kudoh, T. Tanaka, Phys. Rev. D{\bf 68}, 044025
(2003) [gr-qc/0305006].


\bibitem{koyn}
K. Koyama, unpublished notes.


\end{thebibliography}
\end{document}